# Study of Local Nonlinear Properties Using a Near-Field Microwave Microscope

Sheng-Chiang Lee and Steven M. Anlage, *Member, IEEE*

*Abstract*—We have developed a near-field microwave microscope to locally apply microwave frequency currents and fields to superconductors, and dielectric substrates, and measure the locally generated 2$^{nd}$ and 3$^{rd}$ harmonic responses. We measure the local nonlinear response of a $Tl_2Ba_2CaCu_2O_y$ film grown on an MgO substrate, and observe a large response due to the enhanced current density near the edge. We also study the local nonlinear response of a $YBa_2Cu_3O_{7-\delta}$ thin film grown on a bi-crystal $SrTiO_3$ (STO) substrate, and spatially identify the grain boundary through higher harmonic measurements. The spatial resolution is determined by the magnetic loop probe size. A scaling current density $J_{NL}$ is extracted to quantify the magnitude of the nonlinearity of the superconductor. Preliminary results on the nonlinear properties of some commonly used substrates, e.g. MgO and STO, have also been obtained.

## I. INTRODUCTION

THE nonlinear properties of high-$T_c$ superconductors have been of great concern in microwave applications due to their undesired intermodulation response at moderate power levels. While all superconductors have an intrinsic nonlinearity associated with the nonlinear Meissner effect, extrinsic nonlinearities due to topographic features, structural defects, and edge-current buildup are also inevitable in sample fabrication and applications. Several microscopic models, including BCS theory [1], [2], Ginzburg-Landau theory [3], and microwave field-induced modulation of the super/normal fluid density near $T_c$ [4], have been proposed to understand the intrinsic nonlinearity of superconductors. However, upon comparing the expected intrinsic nonlinearity from these models with the extrinsic nonlinearity, the extrinsic nonlinear sources are seen to dominate the nonlinear behavior of many superconductors at most temperatures below $T_c$ [5]. Many experiments have studied the nonlinearity of superconductors in terms of intermodulation distortion [5], [6], harmonic generation [7], [8], or the nonlinear surface impedance [9]-[11]. However, most of these experiments are done with resonant techniques, which by their nature study the averaged nonlinear response from the sample rather than locally, hence have difficulty in either avoiding edge effects, or determining the homogeneity, in terms of nonlinearities, of the sample. Therefore, a technique capable of locally measuring nonlinear properties of a sample is necessary for understanding the microscopic origins of superconducting nonlinearities.

## II. EXPERIMENT AND DATA

### A. Experimental Setup

In prior work [6], we studied the intermodulation signal from a high-$T_c$ superconducting microwave resonator using a scanning electric field pick-up probe. Both the "global", and the "local" intermodulation power measured with the open-ended coaxial probe, were presented. However, the local measurements were actually a superposition of nonlinear responses that were generated locally but propagated throughout the microstrip and formed a resonant standing-wave pattern. To avoid this loss of local information, we have developed a non-resonant near-field microwave microscope, to non-destructively measure the local harmonic generation from un-patterned samples. This technique also works equally well through $T_c$ and into the normal state of the sample. It also permits measurements with variable spatial resolution and rf current orientation. Finally, the experiment determines both the second and third harmonic nonlinearities at the same location, and does so at any frequency or temperature of interest.

In our experiment (Fig. 1), low pass filters are used to filter out higher harmonics generated by the microwave source, and guarantee that only the selected fundamental frequency power $P_f$ is sent to the sample, while the reflected harmonics are selected by 2 high pass filters, before being amplified by

Sheng-Chiang Lee is with the Center for Superconductivity Research in the Physics Department of the University of Maryland, College Park, MD 20742-4111 USA (phone: 301-405-0474; fax: 301-405-3779; e-mail: sycamore@wam.umd.edu).

Steven M. Anlage is with the Center for Superconductivity Research in the Physics Department of the University of Maryland, College Park, MD 20742-4111 USA (e-mail: anlage@squid.umd.edu).
This work is supported by DARPA DSO Contract # MDA972-00-C-0010 through a subcontract by STI, the University of Maryland/Rutgers NSF MRSEC under grant DMR-00-80008 through the Microwave Microscope SEF, and NSF IMR under grant DMR-98-02756.

~65dB, and measured by a spectrum analyzer as $P_{2f}$ and $P_{3f}$.

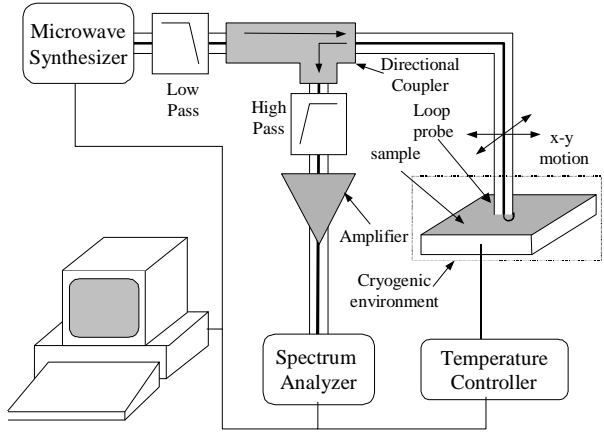

Fig. 1. Experimental Setup. A synthesized tone at approximately 6.5 GHz is first low-pass filtered, and then sent to the tip of the microwave probe. There an rf current is induced in the sample, creating second and third-order nonlinear signals. These signals are gathered by the probe tip and high-pass filtered and amplified before being measured by the spectrum analyzer.

A loop probe, which is made of a coaxial cable with its inner conductor forming a semi-circular loop shorted with the outer conductor, [12] is designed to directionally enhance the magnetic coupling between the probe and sample, and induce microwave frequency currents of a controlled geometry in the sample (Fig. 2), while both the probe and the sample are kept in a high vacuum, cryogenic environment. The harmonic signals generated in the sample couple back to the loop probe to be filtered and measured by a spectrum analyzer. Since the probe can couple to the sample at points far away from the sample edge, the nonlinearity due to the edge-current buildup effect is eliminated. The probe can be translated over the surface of the sample in the x-y plane manually.

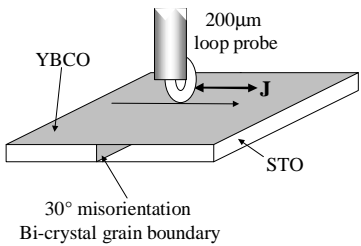

Fig. 2. The setup of a magnetic loop probe coupling to a YBCO/bi-crystal STO sample. The microwave current is determined by the loop orientation, which is put perpendicular to the 30° tilt mis-orientated YBCO grain boundary on the film.

### B. Sample

The superconducting samples we study are a ~5000 Å thick $Tl_2Ba_2CaCu_2O_y$ (TBCCO) film grown on a MgO substrate, and 500 Å thick $YBa_2Cu_3O_{7-\delta}$ (YBCO) thin films deposited by pulsed laser deposition on a bi-crystal STO substrate with a 30° tilt mis-orientation. The distance between the loop probe and the sample is fixed by a 12.5 μm thick Teflon sheet. The $T_c$ of the TBCCO film is 105.6 K, as determined by 4-point zero resistance measurement, and the $T_c$ of the plain YBCO and bi-crystal YBCO are 88.9 K and 88.8 K, respectively, as measured by ac susceptibility. Two commonly-used dielectric substrates, STO (5 mm x 5 mm x 0.5 mm) and MgO (10 mm x 10 mm x 1 mm), are also evaluated by this technique.

### C. Data

Line-cut measurements of the 3rd order harmonic generation are performed on TBCCO using a magnetic loop probe at 102.5 K < $T_c$. While the probe is moved perpendicular to the uni-directionally induced current on the film, a clear increase in 3rd order nonlinear response is observed near the edge of the film, as shown in Fig. 3. The enhanced $P_{3f}$ near the edge is due to the large screening current at the edge required to prevent magnetic field, generated by the microwave current, from penetrating into the superconductor. All other measurements presented below take place in the middle of the film where the edge effects are not significant.

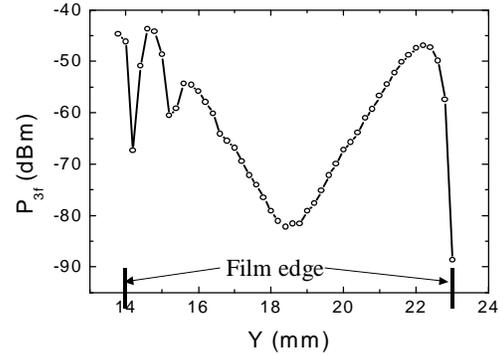

Fig. 3. Line cut of $P_{3f}$ along the direction perpendicular to the induced microwave current on the TBCCO film taken at T = 102.5K. A 6.5GHz microwave signal of +10dBm is applied to the sample. $P_{3f}$ is enhanced tremendously upon approaching both edges, located at ~Y=14mm, and 23mm.

Measurements of the temperature dependent 2nd and 3rd order harmonic power ($P_{2f}$ and $P_{3f}$) are performed on the YBCO bi-crystal film, both above the grain boundary (GB) and far away from the grain boundary (non-GB). A strong peak in $P_{3f}$ (T) near $T_c$ is observed at both locations. [18] However, at temperatures far below $T_c$, $P_{2f}(T)$ and $P_{3f}(T)$ show strong temperature dependence above the GB, while they are not distinguishable from the noise in the non-GB areas. While Josephson vortices should be responsible for the observed $P_{2f}$, [13], [14] $P_{3f}$ is due to the nonlinear impedance of the Josephson junction. [15]

Such a clear distinction between the response of GB and non-GB areas enables us to use this microscope to locally identify the bi-crystal grain boundary as a nonlinearity source in superconductors (Fig. 4). Here the probe is translated over the GB, which shows up as an enhanced nonlinear response. The spatial resolution is determined by the loop probe size, which is ~500 μm in the experiment.

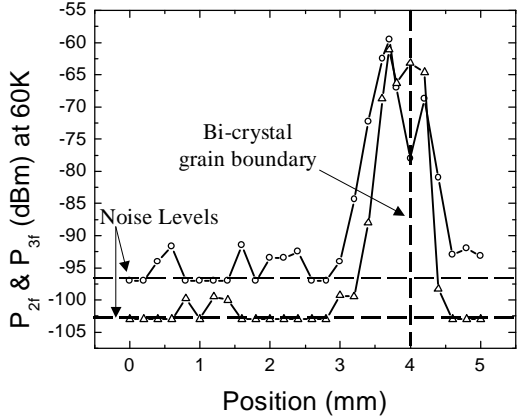

Fig. 4. Measurements of $P_{2f}$ and $P_{3f}$ versus position perpendicular to a YBCO bi-crystal GB at 60K, while a +8 dBm, 6.5 GHz microwave signal is applied through a magnetic loop probe. The grain boundary is located near a position of 4 mm. Circles denote the 2nd and triangles denote the 3rd harmonic powers measured with a spectrum analyzer.

To further evaluate the capability of the microscope to detect superconducting nonlinearities due to different mechanisms, a scaling current density $J_{NL}$ is extracted from the data in Fig. 4 based on the assumption of a leading-order quadratic dependence of the superfluid density on rf current density: $\lambda_L(T,J) \cong \lambda_L(T,0)(1+(J/J_{NL}(T))^2)^{1/2}$,
where $\lambda_L$ is the London penetration depth, and $J_{NL}$ is the scaling current density, whose magnitude is determined by the nonlinearity mechanism. [5], [16]

We calculated numerically the current distribution on the superconducting film according to our experimental geometry, and use it to estimate the nonlinear inductance of the superconductor. Following Booth's algorithm [16], the conversion between $J_{NL}$ and measured $P_{3f}$ is established. Fig. 5 shows the converted $J_{NL}$ from the $P_{3f}$ data in Fig. 4. The $J_{NL}(x)$ distribution shows a similar spatial resolution to both the $P_{3f}$ and the probe surface current distribution. From the magnitude of the $J_{NL}$ we extract (~ $1.5 \times 10^5$ A/cm$^2$), and its location, we confirm that the nonlinear response we observe is due to the bi-crystal grain boundary that forms a long Josephson junction along the boundary. Comparing the extracted $J_{NL}$ with the critical current density $J_c$ of YBCO bi-crystal junction at 30°, [17] we find that the $J_{NL}$ is comparable to the $J_c \sim 10^4$-$10^5$ A/cm$^2$ of the YBCO junction. It is also worth noting that the $J_{NL}$ extracted for the Josephson nonlinearity appears to be power dependent. This is because $J_{NL}$ is calculated on the assumption of a quadratic power-law nonlinearity, and the Josephson nonlinearity is generally not a quadratic power-law[5]. Nevertheless, the calculation of $J_{NL}$ still gives a quantitative understanding of the magnitudes of different nonlinear sources.

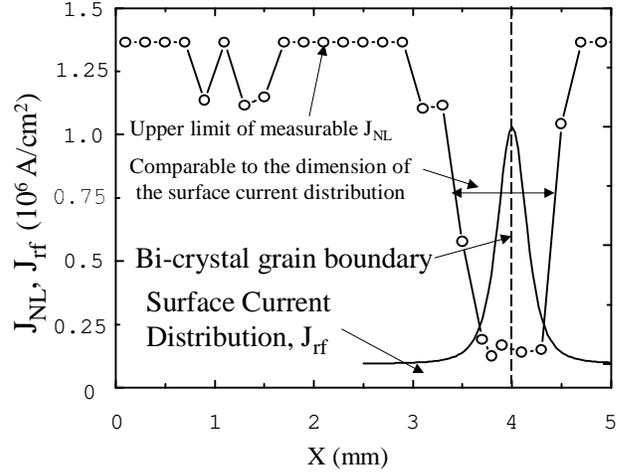

Fig. 5. The calculated nonlinear current scale, $J_{NL}$, from the $P_{3f}$ data in Fig. 4. The upper limit corresponds to the noise level in our current setup, which is about -103dBm. The curve plotted on the bottom is the calculated current density distribution beneath the probe, whose maximum is ~ $10^6$ A/cm$^2$.

The sensitivity of our current setup is limited to $J_{NL} \leq 1.3 \times 10^6$ A/cm$^2$. Since the magnitudes of the measured nonlinear responses are determined by the loop-sample coupling, enhancing this coupling will increase our sensitivity to $J_{NL}$. This can be done by reducing the size of the probe, so that the effective distance between the probe and sample is reduced, and the coupling enhanced. Our calculations also suggests that as long as the film thickness is less than the penetration depth, thinner films will generate stronger nonlinear responses from the same nonlinear source (same $J_{NL}$) while applying microwaves of the same power. More discussion of our work on this YBCO bi-crystal grain boundary with Josephson junction model and $J_{NL}$ can be found in Ref. [18].

It has recently become clear that the dielectric substrates used in high temperature superconducting microwave devices are also nonlinear [19], [20]. Therefore, similar experiments of 2$^{nd}$ and 3$^{rd}$ harmonic generation were performed on bare MgO and STO substrates using the magnetic loop probe.

A clear temperature-dependent $P_{2f}(T)$ is observed on MgO (Fig. 6), while $P_{3f}(T)$ is not detectable above the noise with the loop probe between 4 K and 110 K. A defect dipole relaxation model has been suggested to interpret the measured nonlinear properties of MgO. [18] The second harmonic signal we see here is consistent with the Fe-impurity mechanism proposed as the source of nonlinearity in MgO [18]. Our experiment may provide another approach to study the nonlinear response of MgO, although the data is preliminary at this moment.

Temperature dependent $P_{2f}(T)$ and $P_{3f}(T)$ are also observed on a STO substrate, within narrow temperature windows at certain temperatures (Fig. 7). We believe this is due to its strongly temperature-dependent dielectric constant, so that when a fixed-frequency microwave signal is applied, the substrate becomes a microwave dielectric resonator at temperatures corresponding to the appropriate dielectric constants. The enhanced $P_{2f}$ and $P_{3f}$ may be due in part to resonantly-enhanced background nonlinearity signals.

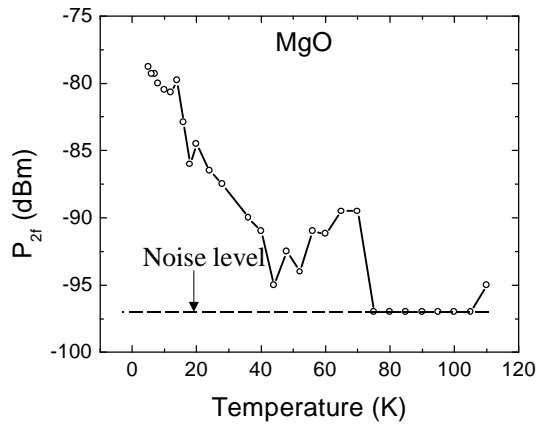

Fig. 6. Temperature dependence of $P_{2f}$ of an MgO substrate, measured by a magnetic loop probe with a primary tone at 6.5GHz.

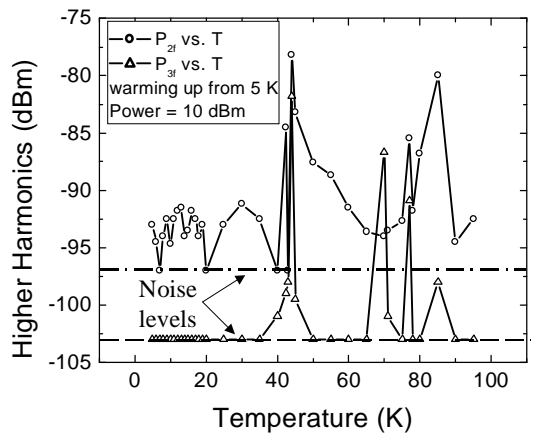

Fig. 7. Temperature dependent $P_{2f}(T)$ and $P_{3f}(T)$ of STO, measured by a magnetic loop probe with a primary tone at 6.5GHz. The dashed and dot-dashed lines are the noise levels for $P_{2f}$ (upper) and $P_{3f}$ (lower) measurements, respectively.

## III. CONCLUSION

We have shown that our near-field microwave microscope is capable of identifying local nonlinear sources, such as a bi-crystal grain boundary and edge-effects, with a spatial resolution similar to the loop probe size. By reducing the loop probe size, the spatial resolution can be further improved. We are also able to convert the measured $P_{3f}$ to a scaling current density $J_{NL}$, which provides a common ground to quantitatively compare the magnitudes of different nonlinearities. The sensitivity of our microscope to $J_{NL}$ can be improved by reducing the probe size and the film thickness. On the other hand, our preliminary work on nonlinear dielectrics, such as MgO and STO, also shows the capability of using this microscope to locally characterize nonlinear dielectric substrates, which is helpful in evaluating the impact of such substrates in microwave applications of superconducting devices.

## ACKNOWLEDGMENT

We thank STI for making the TBCCO film, Su-Young Lee for making the YBCO thin films, and Greg Ruchti for HFSS calculations.